# The Donation-Payment Gift Card Concept: how to give twice with one card

R. Crane[1], J. Escobar[1] and D. Sornette[2,3]

[1] Department of Physics and Astronomy
University of California, Los Angeles, California 90095
[2] Institute of Geophysics and Planetary Physics
and Department of Earth and Space Science
University of California, Los Angeles, California 90095.
[3] Laboratoire de Physique de la Matière Condensée
CNRS UMR6622 and Université des Sciences, B.P. 70, Parc Valrose
06108 Nice Cedex 2, France
E-mails: rcrane@physics.ucla.edu, escobar@physics.ucla.edu,
sornette@moho.ess.ucla.edu

***Abstract.***
Currently, there are large amounts of unused funds associated with Pre-paid Stored Value cards. These funds are being claimed by either the companies that issue the cards, or State Governments. This has given rise to many complicated legal issues whose outcome does not remove the basic fact that the consumer has no power of decision over the final destination of these funds. We propose a Donation-Payment card that automatically donates all unused funds to the consumer's preferred charity, which reinstates their choice and resonates with the innate altruistic human desire. The partnership with charities should increase significantly the number of people who use this Pre-paid Stored Value card, ensuring its development as the fastest growing new payment instrument to arrive since the introduction of debit cards.

**Introduction**

Standard economic theory, starting with Adam Smith's invisible hand, holds that those who trade for their own selfish motives of maximizing their private preferences may contribute more to the public wealth than those who claim altruistic motives. Under restrictive conditions, this has been shown to result from a self-organizing mechanism acting at the global system level, which ensures that optimal allocation of resources derives from competition within the rule of law. In a nutshell, according to this view, being competitively selfish adds value for all.

But to what degree can doing good add competitive value to products and services?

Many companies have already discovered the benefits of offering products and services that appeal to the altruistic nature of consumers. Famously, Henry Ford made his car affordable to the mass, paid wages twice the going rate (and was sued for this by outraged stockholders) and developed an image of high quality and more responsible product for which eager consumers willingly paid for. As a result, he made enormous profits, the industry moved to his standards, and the automobile was transformed from a luxury to an affordable staple.

In some cases winning the sympathy of consumers outweighs raising their profit margin; especially if by this, one captures a large enough market. A growing body of evidence has indeed confirmed that consumer's preferences are not based on pure monetary or wealth utility. There are many examples showing that measures of happiness become weakly linked with income or wealth above a threshold level and that people are often driven by non-profit motives, such as altruism. This important fact has been recognized by many top corporations and has been incorporated as a branch of their marketing strategies, in what is referred to as 'cause related marketing'.

As an example, 'Working Assets', a telecommunications and credit card company, directs one percent of their total income to a variety of progressive non-profit organizations. Following this philosophy, they have not only raised over $47 million since 1985, but they have also taken over a sizable fraction of the market. This model works mainly because the donation process is automatic and satisfies the consumer's desire to give back to society.

But donating money through purchases is not the only effortless way in which small contributions can accumulate into large, substantial amounts.

The *Unicef Change for Good program*, has shown that it is possible for charities to benefit greatly from leftover money. This program collects unused foreign coins and notes from airplane passengers, and has raised $37 million dollars over a period of 14 years. The success of this program relies both on the effortless donation process and on the access to sources of unused money. A very similar source of unused money exists in the gift card market, a market that has enormous potential for new e-commerce payment methods.

In what follows, we will detail how this leftover source of money can be channeled to charities in a way that elegantly solves legal issues while providing the best free market solution for the stored value prepaid card industry.

**The State of the Payment Industry**

The growth of payment instruments beyond the primitive forms of barter has co-evolved with the development of commerce and of civilizations from historical times to present. Rapid changes in the use of payment instruments are occurring around the world, evidenced by the decline in check usage in the United States, the growth of debit and credit card payments in many countries, the development of personal online payment methods, the redesign of large-value payment systems in many countries, and the response to antitrust disputes involving card-based payments. Money and, more generally, payment instruments have more than just the purely economic, narrow and technical functions usually attributed to them but have widely recognized social, institutional and psychological aspects. The gift card industry is a particularly vivid example of such a social phenomenon. Nearly one-half of U.S. consumers used a gift card during the past year (2004) according to results of Standard Register's third National Consumer and Retailer Survey of Plastic Card Usage. Polls indicate that 45 percent of adults have used gift cards, an increase of 34 percent over the previous year's survey.

Recently, Stored-Value Prepaid Cards (also known as "branded" or "open-system" gift cards and henceforth referred to as SVP-Card) carrying the logo of Visa, Mastercard, and American Express have been introduced into circulation at grocery and retail outlet stores. These disposable, preloaded payment cards function in much the same way as a debit-card, except that they are not tied directly to any account, and are anonymous (not being associated with any user). They have risen in popularity because they are easily obtainable and provide the user with anonymity and the security desired for small, online transactions (music, books, subscriptions, etc.).

**Legal Limbo: Whose Money is it?**

An apparently minor issue is however becoming a major concern for the pre-payment card industry: how to deal with the unused value of each card?[1] It is estimated that anywhere from five to 14 percent of the value of each card goes unused (on average). With somewhere between seven and 20 million cards of this type issued in 2004 at an average value of $50, this adds up to an enormous amount estimated between $19 million to $140 million, (expected to grow to 50 million cards by 2010). Recent research in social psychology and behavioral finance suggest that the phenomenon of unused value in prepaid cards is not going to disappear, simply because of human "rational inattention." Recently, lawsuits have been brought against several companies by the states for overstepping their legal claim to this unused money. The States Governments argue that the unused funds qualify as "abandoned property", and as such must be rendered to the State. The card-issuing companies argue that the cost to return these funds to the State is prohibitively expensive, and they must therefore impose fees reducing the card's balance to zero, so that there is no "abandoned property" to redistribute. But it is the imposition of these fees that is forbidden by most state laws regulating SVP-Cards. The alternative of returning the unused funds to the consumer would not only be economically unfeasible, but is not even a possibility because of the anonymous nature of this product.

In Summary:

| | |
|---|---|
| 1 | It is the innate "rational inattention" of consumers that creates a large pool of money in the SVP-Card industry. |
| 2 | The legal status of this money remains unclear. If the status-quo doesn't change, the money will either end up with the state or with the companies. Alternatively this could legally undermine the basis of the SVP-Card industry, removing this payment instrument from the market. In this sense, the unused money is, as a byproduct, a nuisance. |

---

[1] Unused Funds are funds which are not used within some pre-specified time period, usually six months to one year.

3             In effect, this money belongs to the people only collectively since it can not be claimed individually.

**Power to the People:  A Charitable Solution to the Legal and Moral Issues**

From both a legal and moral point of view, the best solution is that it is the consumer who decides the fate of their unused funds.  But the consumer loses their power of decision once the money is pooled collectively.  The solution is therefore to allow the consumer to decide a priori where to channel their unused funds.

We propose that the leftover funds be donated to charity. In this way, the core of the legal conflict evaporates since it is the people who consciously decide where these unspent funds are directed.  This solution resonates with the altruistic nature of humans through focusing resources back into the community.

As our solution to the problem of companies operating in a way which benefits the collective, charitable organizations become the natural recipient and administrator of these funds.  As others have recognized, charities carry advantages over the state in providing social goods since they are more efficient and diverse, as a result of being optimized by constant scrutiny.  Charitable associations are arguably the only entities with the transparency, constant scrutiny and moral character to bring this enormous reservoir of unused monies back to the community.

This approach is not only the optimal solution concerning the SVP-Cards, but also helps create a better social environment by simplifying the donation process.

**Examples and Variations**

As an example of the implementation of this solution, several charities could issue these cards with each card bearing the logo of one specific charity. The consumer would then decide which of these cards to purchase (such as the American Red Cross, American Cancer Society, or Save the Children), thus deciding where their unused funds, if any remained, would be directed.

As pointed out in the introduction, the Donation-Payment card could also be issued by a company. This company would profit only from the up-front fee the consumer pays to purchase the card, and all unused funds would be directed to the company's partnering charities. This would give the issuing-company an edge over its competitors since American's prefer to support companies who give back to society. It is widely acknowledged that a large fraction (60 percent in 2003) of Americans have planned on purchasing a product in which a percentage of the price was donated to a cause. The Donation-Payment card thus kills two birds with one stone by marrying business interests with societal benefits.

Furthermore, the previous estimation of $19 to $145 million dollars per year channeled to the charities is likely conservative, since the partnership with charities would increase the number of people who would use the Donation-Payment card. Think for instance of the state of mind of a grand-parent giving a SVP-card to a grand-child, who feels like "killing two birds with one stone" or should we say more aptly feels like "giving twice with one card".

**Conclusion**

We have proposed the creation of the Donation-Payment card as a solution to the legal and moral issues plaguing the emerging anonymous, direct-to-consumer, stored-value card market. Our proposal to redirect millions of dollars whose status is in conflict not only improves society and empowers people, but also represents the best free-market solution to the SVP-Cards industry.

*Riley Crane is a graduate student in Condensed Matter Physics at the University of California, Los Angeles.*

*Juan Escobar-Sotomayor is a graduate student in the Physics of Energy Focusing Phenomena at the University of California, Los Angeles as a Fulbright scholar.*

*Didier Sornette is Professor at the University of California, Los Angeles, and a research director at Centre National de la Recherche Scientifique, France. He is the author of several books including Critical Phenomena in Natural Sciences (Springer-Verlag, 2000, 2$_{nd}$ ed. 2004), Why Stock Markets Crash: Critical Events in Complex Financial Systems (Princeton University Press, 2003), co-author (with Y. Malevergne) of the Extreme Financial Risks (From dependence to risk management), (Springer, Heidelberg, 2005) and has authored or coauthored more than 330 papers in international journals. He has consulted widely for aerospace companies, banks, investment and reinsurance companies.*

**References and Further Reading**